\documentstyle[12pt,aps,prc,preprint]{revtex}
\begin{document}

\begin{titlepage}
\title{\vspace*{35mm}\bf Constraints on $\Sigma$ Nucleus Dynamics from Dirac
                         Phenomenology of $\Sigma^-$ Atoms}
\vspace{6pt}

\author{J.~Mare\v{s}$^{a,b}$, E.~Friedman$^{a,c}$, A.~Gal$^c$
        and B.K.~Jennings$^a$ \vspace{6pt} \\
\it $^a$ TRIUMF, 4004 Wesbrook Mall, Vancouver, B.C., Canada V6T 2A3\\
$^b$ Nuclear Physics Institute, 25068 \v{R}e\v{z}, Czech Republic \\
$^c$ Racah Institute of Physics, The Hebrew University, Jerusalem 91904,
Israel}

\vspace{6pt}
\maketitle

\begin{abstract}
Strong interaction level shifts and widths in $\Sigma^-$ atoms are analyzed
by using a $\Sigma$ nucleus optical
potential constructed within the relativistic mean field approach. The analysis
leads to potentials with a repulsive real part in the nuclear
interior. The data are sufficient to establish the size of the isovector
meson--hyperon coupling. Implications to $\Sigma$ hypernuclei are discussed.
\end{abstract}\vspace{1cm}
\centerline{\today}
\end{titlepage}

\section{Introduction}

The relativistic mean field (RMF) approximation [1] is an interesting
alternative to more conventional approaches to the nuclear many-body problem.
Numerous studies have demonstrated the usefulness of the RMF approach in
describing the behavior of nucleons in the nuclear medium. Recent RMF
calculations of $\Lambda$- and multi $\Lambda$ hypernuclei [2-4] have shown
that it can be successfully extended to the more general baryon-nucleus systems
[5-9]. Hyperon-nucleus scattering has been studied as well [10]. In these
works, the meson--hyperon coupling constants were either fitted to known
hypernuclear spectra or obtained from the parameters of a nucleon shell model
by
using the constituent quark model.

Calculations of $\Sigma$-hypernuclei [11] have often been based on analyses
of $\Sigma^-$ atomic data [12,13] in terms of attractive and absorptive
$\Sigma$-nucleus potentials, yielding in the nuclear interior depth values
\begin{eqnarray}
 -{\rm Re} V^{\Sigma}_{opt}(0) & \approx &  \mbox{25--30~MeV, }
 \\
 -{\rm Im} V^{\Sigma}_{opt}(0) & \approx &  \mbox{10--15~MeV}.
\end{eqnarray}
However, a considerably shallower potential, particularly for
Re$V^{\Sigma}_{opt}$, is indicated by fitting to $\Sigma$-hypernuclear spectra
from ($K^-$,$\pi^+$) reactions [14]. Furthermore, a very recent
phenomenological
analysis of level shifts and widths in $\Sigma^-$ atoms
by Batty {\it et al}. [15] suggests that Re$V^{\Sigma}_{opt}$ is
attractive only at the nuclear surface, changing into a repulsive potential
as the density increases in the nuclear interior. Although the magnitude of the
repulsive component cannot be determined unambiguously by the atomic data,
the small attractive pocket of such potentials does not provide sufficient
binding to form $\Sigma$-hypernuclei. This conclusion has been independently
arrived at by Harada [16], folding a soft-core repulsive plus long-range
attractive $\Sigma N$ interaction into the nuclear density.

In view of the above seemingly conflicting observations, it is topical to
use the RMF approach directly for constructing the $\Sigma$-nucleus optical
potential by fitting to $\Sigma^-$ atomic data. The aims of the present
work are:
\begin{enumerate}
  \renewcommand\theenumi{\roman{enumi}}
  \renewcommand\labelenumi{\theenumi)}
\item to check whether or not the RMF approach can reproduce the atomic data;
\item to establish constraints imposed by fitting to the data on the scalar
and vector meson couplings to a $\Sigma$ hyperon;
\item to establish the size of the isovector-vector meson-hyperon coupling;
\item to investigate the sign of the $\Sigma$-nucleus potential in the nuclear
interior and whether or not $\Sigma$ hypernuclear bound states are expected to
exist.
\end{enumerate}

The paper is organized as follows. Sect.~2 presents updated fits of a
phenomenological density dependent (DD) optical potential.
In sect.~3 we introduce the RMF optical potential for $\Sigma$, discuss
the constraints imposed by the atomic data on its form and present results.
These RMF results are compared with results of other approaches in sect.~4
which also offers a discussion and summary of our conclusions.

\section{Phenomenological density dependent optical potential}

Before comparing the $\Sigma$-nucleus potentials based on the RMF
approach to those derived using the phenomenological analysis, we took the
intermediate step of replacing the ``macroscopic" nuclear densities (such as
2-parameter Fermi distributions) used in the previous phenomenological analysis
[15] by more realistic ``microscopic" densities.  The motivation for this
replacement is twofold: first, we wish to use densities that are more
consistent with the resulting RMF densities, and second, it is desirable to
use densities that are as appropriate as possible outside the nuclear surface,
for discussing $\Sigma^-$ atoms.

In order to appreciate the latter point, we have repeated the ``notch test''
of ref. [15] in a modified form to see which range of radii has a significant
effect on the atomic levels.  The sensitivity of the data to the potential
around the radius $R_N$ is obtained by multiplying the best-fit potential
by a factor:
\begin{equation}
f=1-d \; exp[-(\frac{r-R_N}{a_N})^2 ]
\end{equation}
representing a ``notch'' in the potential around $ r=R_N $ spread approximately
over $\pm a_N$ (in fm), whose relative depth is $d$. By varying the depth $d$
and observing the changes in $\chi^2$, the sensitivity (defined as the values
of
$d$ that cause $\chi^2$ to increase by one unit) can be determined. For $a_N$
the value of 0.5 fm was chosen.  By scanning over $R_N$, the radial region
where the fit to the data is sensitive to the potential is determined.  Figure
1 shows the results obtained, using ``microscopic" densities as described
below,
by scanning $R_N$ in steps of a nuclear diffuseness $a_0$, i.e.,
$R_N=R_0+a_0 \Delta$, with $R_0=1.1A^{1/3}$ fm and $a_0=0.5$ fm.
It is seen that the radial region which is ``sampled'' by the data
is outside of the nuclear surface. This updates the similar results shown in
Fig.~3 of ref. [15] using ``macroscopic" densities.

 The method chosen to generate ``microscopic" nuclear densities was to fill in
single particle (SP) levels in Woods-Saxon potentials separately for
protons and neutrons. The radius parameter of the potential was adjusted
 to reproduce the {\it rms} radius of the charge distribution
(after folding in the finite size of the proton). The binding energy of the
least bound particle was set equal to the corresponding separation energy.
For $N>Z$ nuclei the {\it rms} radius of the neutron density
distribution was chosen to be slightly larger than that of the protons.
By using this method, rather realistic density distributions are obtained
in the region of the exponential fall-off. These densities are less
reliable in the nuclear interior, but the $\Sigma^-$ atom data are not
sensitive to this region as is clearly seen from Fig.~1.

Several of the fits of ref. [15] have been repeated using these SP
densities. The $\Sigma^-$ nucleus potential is written as

\begin{equation}
2\mu V^{\Sigma}_{opt}(r) =
 -4\pi(1+\frac{\mu}{m})\left\{ \left[b_0+B_0(\frac{\rho(r)}{\rho(0)})^\alpha
\right]
\rho(r)+\left[b_1+B_1(\frac{\rho(r)}{\rho(0)})^\alpha\right]\delta\rho(r)
\right\}
\end{equation}
where $\mu$ is the $\Sigma$-nucleus reduced mass, m is the mass of the nucleon
and $\rho(r)=\rho_n(r)+\rho_p(r)$ is the nuclear density distribution
normalized to the number of nucleons A. The isovector density distribution
is given by $\delta\rho(r)=\rho_n(r)-\rho_p(r)$. The parameters $b$ and $B$
are given in units of fm and are obtained from fits to the $\Sigma^-$
atom data.

For the density-independent $t_{eff}\rho$ potential ($B_0=0, B_1=0$), the
results obtained with the SP densities were very similar to those obtained
before [15], namely, values of $\chi^2$ per degree of freedom, $\chi^2/F$,
were about 2.4 and the isovector part of the potential was not determined
by the data. With the introduction of density dependence into the
potential, substantial improvement in the fit to the data is observed.
The best-fit values of $\chi^2$  obtained with SP densities are
significantly lower than those obtained with the schematic parameterization
of the nuclear densities [15]. Moreover, the real part of the isovector
potential is now found to be reasonably well-determined, presumably because
the SP densities represent  more reliably the differences between protons
and neutrons outside of the nuclear surface.

Results of two DD fits are summarized in table 1.  Potential I is the analog of
potential D or A' of ref. [15], where the density dependence is introduced only
into the real isoscalar part via the parameter Re$B_0$. It is assumed
throughout that Im$b_1$=$-$Im$b_0$ so that $\Sigma^-$ particles are absorbed
only on protons.
All four parameters varied in the fit are phenomenological and it is seen that
they are reasonably well-determined.  The DD exponent $\alpha$ was held
fixed during the parameter search and its value was subsequently varied between
0.1 and 1.1 and fits repeated.  Potential II is analogous to the potential
of ref. [15] for which the linear density terms were held fixed at values
suggested by One Boson Exchange (OBE) models: $b_0$=1.2$+i$0.45 fm,
$b_1$=$-$0.45$-i$0.45 fm. In this case, three adjustable
parameters, namely, the complex $B_0$ and Re$B_1$ (Im$B_1$=$-$Im$B_0$), were
obtained from fits to the data. In both cases the fits are excellent and in
both cases the variation of $\chi^2$ with $\alpha$ was weak. The values shown
in table 1 are typical examples and $\alpha$ could not be determined uniquely
in the range of 0.4 to 0.9.

{}From the parameter values in table 1 it is seen that whereas the potentials
are
attractive at low densities, they become repulsive as the density approaches
nuclear matter densities, a feature that had already been observed [15]. Figure
2 shows the real and imaginary potentials for Si, Ca and Pb near the nuclear
surface. It is seen that there is a
small attractive pocket near the surface and that the real potential becomes
repulsive toward the interior. The imaginary potential is purely absorptive;
similar results for Re$V_{opt}^\Sigma$ are obtained if Im$V_{opt}^\Sigma$
is made to saturate at nuclear matter density [17]. It should be mentioned
that with the limited data available and the very different quality of various
data, the isovector part of the potential is determined solely by the
experimental results for $\Sigma^-$ atomic Pb.

\section{Construction of the RMF optical potential}

The RMF formalism describes baryons as Dirac spinors coupled to scalar
($\sigma$) and vector ($\omega$, $\rho$) meson fields. The underlying RMF model
used here is based on the Lagrangian density of the form:
\vspace{5mm}
\begin{equation}
\hspace{-20mm}{\cal L} = {\cal L}_{N} + {\cal L}_{\Sigma} \;\; ,\;\;\;\;
\end{equation}
\begin{equation}
\hspace{-20mm}{\cal L}_{\Sigma} ={\bar \Psi}_{\Sigma} \left[ \; i\,\gamma_{\mu}
\partial^{\mu} - g_{\omega \Sigma}\, \gamma_{\mu} \omega^{\mu} -
\left(M_{\Sigma}
 + g_{\sigma \Sigma}\sigma \right)\right] \Psi_{\Sigma}
 + {\cal L}_{\rho \Sigma} + {\cal L}_{A \Sigma} \; \; \; ,
\end{equation}
\begin{equation}
{\cal L}_{\rho \Sigma} + {\cal L}_{A \Sigma} =
-{\bar \Sigma}_{ij}\left( \; {g_{\rho \Sigma} \over 2} \gamma_{\mu}
\Theta^{\mu}_{jk} + {e \over 2}\gamma_{\mu}A^{\mu}
({\tau}_3)_{jk}\right)
\Sigma_{ki}\;\;\;\; ,
\end{equation}
where
\begin{equation}
\hspace*{-10mm}  \Sigma =
\left( \begin{array}{cc}
\Psi_{\Sigma^{0}} & \sqrt{2}\Psi_{\Sigma^{+}} \\
\sqrt{2}\Psi_{\Sigma^{-}} & -\Psi_{\Sigma^{0}} \\
\end{array} \right) \;\;, \;\;\;\;\;
\tau_3=\left( \begin{array}{cc}
 1 & 0 \\
 0 & -1 \\
\end{array} \right) \;\;\;,
\end{equation}
and
\begin{equation}
\Theta^{\mu} =
\left( \begin{array}{cc}
\rho_{0}^{\mu} & \sqrt{2}\rho_{+}^{\mu} \\
\sqrt{2}\rho_{-}^{\mu} & -\rho_{0}^{\mu} \\
\end{array} \right)\;\;\;\;.
\end{equation}

The standard form of ${\cal L}_{N}$ can be found elsewhere [1].  The
Lagrangian density ${\cal L}_{\Sigma}$ includes interactions of a $\Sigma$
hyperon with the isoscalar ($\sigma$, $\omega$) and isovector ($\rho$) meson
fields, as well as with the photon generating in leading order the
charged $\Sigma$ Coulomb field due to the nuclear charge distribution.
The Euler-Lagrange equations lead to
a coupled system of equations of motion for both baryon and meson fields.
For the $\Sigma$ particle, and retaining only the timelike component $\mu$ = 0
for the vector fields in the mean (meson) field approximation, the resulting
Dirac equation acquires the form:
\begin{equation}
\hspace{-2mm} [ -i\, \mbox{\boldmath $\alpha$} \cdot \mbox{\boldmath $\nabla$}
 +  V(\mbox{\boldmath $r$}) + \beta (M +
S(\mbox{\boldmath $r$}) )
+ V_{\rm Coulomb}(\mbox{\boldmath $r$})\, \, ]\psi(\mbox{\boldmath $r$}) =
E\psi(\mbox{\boldmath $r$}) \;\;\; ,
\end{equation}
where
\begin{equation}
S(\mbox{\boldmath $r$}) = g_{\sigma \Sigma} \sigma(\mbox{\boldmath $r$})
\;\;\;\;\;
{\rm and}
\;\;\;\;\;
V(\mbox{\boldmath $r$}) = g_{\omega \Sigma}\omega(\mbox{\boldmath $r$})
+ g_{\rho\Sigma}I_3\rho_0(\mbox{\boldmath $r$})
\end{equation}
are the scalar and vector potentials, respectively. In eq. (11),
$I_3 = +1,0,-1$
for $\Sigma^+$, $\Sigma^0$, $\Sigma^-$, respectively.  More details about the
equations of motion and their solution are given in ref. [18].

For each particular nucleus we constructed the Schr\"{o}dinger equivalent (SE)
$\Sigma$-nucleus potential from the scalar (attractive) and vector (repulsive)
Dirac potentials. The SE potential $V_{SE}(r)$ was then used as a real part of
the optical potential in the calculations of $\Sigma^-$ atoms:
\begin{equation}
{\rm Re} V^{\Sigma}_{opt}(r)
\equiv V_{SE}(r) = S(r) + {{E V(r)}\over M_{\Sigma}}
+{(S^2(r)-V^2(r))\over {2M_{\Sigma}}} \;\;\; .
\end{equation}
It is to be stressed here that the isovector part of the real potential comes
out quite naturally in the RMF approach as one allows for coupling the
$\Sigma$ with the $\rho$ field (see eq.(11) for V(r)).

The imaginary part of the $\Sigma$-nucleus optical potential describes the
conversion of $\Sigma^-$ to $\Lambda$ via the reaction
$\Sigma^- p \longrightarrow \Lambda n$. The expression for the imaginary part
of the potential (and consequently the $\Sigma$-hypernuclear width) was
derived by Gal and Dover [13] and Dabrowski and Rozynek [19].
The values of the well-depth, $- {\rm Im} V^{\Sigma}_{opt}(0)$,
were estimated to be between 11.6 and 14~MeV in
nuclear matter. For finite nuclei, a reduction from these estimates should be
expected. In the present work a phenomenological imaginary potential of the
form
\begin{equation}
{\rm Im} V^{\Sigma}_{opt}(r) = t\rho_p(r)\;\;
\end{equation}
was used, where the proton density $\rho_p$ was calculated in the RMF model and
$t$ was taken as a parameter to be determined by fitting the atomic data. The
use of $\rho_p$ in the construction of the imaginary potential reflects the
fact that the $\Sigma^-$ conversion takes place exclusively on protons.

For ${\cal L}_{N}$ the linear (L) parametrization of Horowitz and Serot [18]
was used as well as the nonlinear (NL) parametrization $NL1$ of Reinhard
{\it et al.} [20]. For ${\cal L}_{\Sigma}$ the hyperon couplings were
characterized as in previous works [2-10] via
the coupling constant ratios $\alpha_{i} = {g_{i \Sigma} \over
g_{i N}}$, ($i=\sigma$, $\omega$, $\rho$). In the case of the vector meson
coupling ratio $\alpha_{\omega}$ three values, namely 1/3, 2/3 and 1,
were used. The choice of $\alpha_{\omega}=1/3$ was inspired by a number
of early hypernuclear calculations [2,21-24] in which the strength of a hyperon
(mainly $\Lambda$) coupling used was between 1/3 and 0.4.
On the other hand, the 2/3 ratio follows from the constituent quark model.
Finally, for $\alpha_{\omega}=1$, the equality
$g_{\omega \Sigma}=g_{\omega N}$ is motivated by a recent
QCD sum rule evaluation for $\Sigma$ hyperons in nuclear matter [25].

For each of the above choices of $\alpha_{\omega}$ the ratio
$\alpha_{\sigma}$ together with $t$ from eq. (13) were fitted to the
experimental atomic shift and width in Si. This particular atom was singled
out due to its relatively accurate shift and width data among $N = Z$
(isoscalar) core nuclei. The isovector part of the optical potential was then
included and the ratio $\alpha_{\rho}$ was determined by fitting the
shift and width in Pb while holding the isoscalar parametrization fixed.
In such a way, we found for each RMF model, whether L or NL, three sets of
parameters that give, by construction, an excellent fit for Si and Pb.
Typical results are illustrated in Fig.~3 for $\Sigma$ nucleus NL
optical potentials in Si. When going from $\alpha_{\omega} = 1/3$
to 1, Re$V^{\Sigma}_{opt}$ changes from attraction to repulsion in the
nuclear interior. In addition, the imaginary
part Im$V^{\Sigma}_{opt}$ becomes more absorptive. The same holds for the
linear model which, for a given value of $\alpha_{\omega}$, predicts more
attraction for Re$V^{\Sigma}_{opt}$ and less absorption for
Im$V^{\Sigma}_{opt}$ than the NL parameterization yields, as is shown for
$\alpha_{\omega}=2/3$ in Fig.~4 where $\Sigma$ optical potentials
for Pb are compared with each other. We observed that the $\rho$ - $\Sigma$
coupling is determined unambiguously by the fit to the Pb data, with a value
$\alpha_{\rho} \approx 2/3 $ in all cases.

Having determined the isoscalar as well as isovector parameters of
$V^{\Sigma}_{opt}$ by fitting the Si and Pb data we constructed SE optical
potentials for all nuclei for which $\Sigma^-$ atomic data exist.  Whereas
for $\alpha_{\omega}=2/3$ and 1 we obtained reasonable values of $\chi^2$
without any further adjustments, we failed to get below $\chi^2=26.5$ for
$\alpha_{\omega}=1/3$. The results for the linear model are presented in
Table~2. It is to be noted that we did not aim strictly at the lowest $\chi^2$
value but merely to get $\chi^2 \approx 23$, which is a typical value in
previous DD fits [15].  Consequently, only the value of $\chi^2$ for the
potential L3 from Table~2 represents a true minimum for this particular
parametrization. From this point of view a comparison of $\chi^2$ values listed
in Table~2 might even be too favorable to the L3 potential.

Although the $\Sigma^-$ atomic data are of a poor quality and, moreover,
determine the shape of $V^{\Sigma}_{opt}$ only around the surface and outside
the nucleus as demonstrated by the ``notch'' test (Fig.~1), these data
nevertheless significantly constrain the RMF parametrization of the $\Sigma$
nucleus optical potential. Not only that the value of the $\rho$-$\Sigma$
coupling ratio $\alpha_{\rho} \approx 2/3$ holds unambiguously for all
the parametrizations used, but in addition, a weak $\omega$-$\Sigma$ coupling
is almost certainly ruled out by fitting to the $\Sigma^-$ atom data.
Consequently, the data from $\Sigma^-$ atoms imply, within the RMF approach, a
$\Sigma$-nucleus optical potential with a repulsion in the nuclear interior and
a shallow attractive pocket outside the nuclear surface. For some light
nuclei (Si, S, Mg), in contrast to Fig.~4 for Pb, the linear model with
$\alpha_{\omega}= 2/3$ predicts that the repulsion in the interior goes
back into a second shallow attractive pocket (less than 3.5~MeV) at the origin,
but this obviously cannot be tested by using  $\Sigma^-$ atom data.

\section{Discussion and Summary}

In this work we have improved on the phenomenological
analysis [15] of $\Sigma^-$ atoms in terms of a $\Sigma$
nucleus DD optical potential by using nuclear density
distributions which account more realistically for the
nuclear surface region and outside of it. The earlier work
[15] established that the isoscalar componenet of
$V_{opt}^\Sigma$ changes in the nuclear surface region
from attraction to repulsion as one penetrates the
nuclear interior, and that the isovector component is
undetermined by the data. While the isoscalar component,
in the present work, remains qualitatively unchanged, the
introduction of these more realistic densities has a
pronounced effect on the extraction of the isovector
component of $V_{opt}^\Sigma$, which for $\Sigma^-$ is
reliably determined to be {\it repulsive}. It
should be recalled that this extraction hinges almost
exclusively on the shift and width data for Pb [26] which
are the most accurate among all available $\Sigma^-$ atom
data to date, not just the $N > Z$ data subset. More
precise measurements, particularly for other $N > Z$ atoms,
are highly desirable.

We have also confirmed the results of ref. [15] that
$V_{opt}^\Sigma$ is determined by the $\Sigma^-$~atom
data only outside the nucleus. Nevertheless, the novel
feature of $V_{opt}^\Sigma$ becoming repulsive is found
to occur at a region of space outside the nuclear radius,
still where $V_{opt}^\Sigma$ is determined, even if not
very accurately, by the $\Sigma^-$ atom data. To have a
glimpse into the nuclear interior, a theoretical model is
necessary, and in the present work we have used RMF as a
working hypothesis. Although the RMF approach was
applied before [7, 8], for studying the binding of
$\Sigma$ hypernuclei, these studies were not constrained
by any sound phenomenology since no $\Sigma$ hypernuclear
bound state has ever been clearly established [11], with
the exception perhaps of a $J^\pi = 0^+, I =
1/2~~{^4_\Sigma{\rm He}}$ bound state [27] which is a too
light system to be used as input to RMF calculations.
Furthermore, recent searches for $\Sigma$ hypernuclear
peaks, some of which had been reported on the basis of
limited statistics, have yielded negative result [28,
29]. This leaves $\Sigma^-$ atom data as the sole source
of any possible $\Sigma$ hypernuclear phenomenology.

The application of the RMF calculational scheme to
$\Sigma^-$ atom data in terms of three coupling-constant
ratios ($\alpha_\omega, \alpha_\sigma, \alpha_\rho$) has
been successful in showing that very good quality fits,
reproducing these data, can be made. The larger
$\alpha_\omega$ is (in the range 0 to 1), the better is
the fit. Of these fits the best ones, in the range
$\alpha_\omega~{\underline \sim}~2/3$ to 1, indeed produce
 SE $V_{opt}^\Sigma$ with a volume repulsion in the nuclear
interior. The $\alpha_\omega$ = 1 fit is an excellent one,
reaching almost as low a $\chi^2$ value as those produced
by the completely phenomenological DD optical potential
fits of table 1. The observation that using relatively high
values of $\alpha_\omega$ is distinctly superior to using
relatively low values of $\alpha_\omega$ within the RMF
$\Sigma$-nuclear dynamics is in contrast to the
situation in applying the RMF approach to $\Lambda$
hypernuclei [2, 3, 6]. There, as stressed in ref. [30], a
wider range of values for $\alpha_\omega$ is roughly
equally acceptable, since for any given value of
$\alpha_\omega$ the constraint of a
$\Lambda$-nucleus potential well depth of about 28 MeV
attraction [31] can be satisfied by deriving an
appropriate value for $\alpha_\sigma$. In the absence of
$\Sigma$ hypernuclear data, no analogous constraint is
operative for RMF $\Sigma$-nuclear applications. The
volume repulsion obtained in this work for the isoscalar
component of $V_{opt}^\Sigma$ in fact precludes binding for
$\Sigma$ hypernuclei, at least for those based on core
nuclei with a small neutron excess, $(N-Z)/A \ll 1$.

It is gratifying to conclude that both approaches, the
phenomenological one and the RMF approach, agree with
each other, for a comparable degree of fit, in producing
an isoscalar repulsion in the nuclear interior and, for
$\Sigma^-$, an added isovector repulsion. We stress that
this isovector repulsion, with $\alpha_\rho
{}~{\underline \sim}~2/3$, was derived independently
of the values assumed by the isoscalar coupling constants
ratios. We recall that the common assumption of a pure $F$
coupling for the $\rho$ meson in SU(3), or equivalently of
a universal $\rho$ coupling to isospin, gives $\alpha_\rho
= 1$, a value which is rather different from that
determined by fitting to the $\Sigma^-$ atom data.
However, it is well known that in the OBE approach which
unlike the RMF approach does use SU(3) guidelines for
connecting the $S = -1$ sector to the $S = 0$ sector, the
nuclear isovector single-particle potential $V_1^{(N)}$ is
generated largely by other meson contributions than that
due to a direct $\rho$ exchange [32]. In the RMF approach,
the $\rho$ field is the only agency through which
$V_1^{(N)}$ or $V_1^{(\Sigma)}$ can be generated, so that
the statement $\alpha_\rho~{\underline \sim}~2/3$ is to be
construed just as implying that $V_1^{(\Sigma)}~
{\underline \sim}~(2/3) V_1^{(N)}$, where the
baryon-nucleus isovector potential contribution is defined
by
\begin{equation}
{V_1^{(B)} {\vec t_B} \cdot {\vec
T / A} \quad , \qquad {\vec t_B} = {{{1 \over 2}\vec
\tau} \atop \vec I} \quad {(N) \atop (\Sigma)} \quad ,
}
\end{equation}

\noindent
and $\vec T$ is the nuclear isospin operator with $T_3 =
(Z - N)/2$. Since the nuclear isovector potential close
to zero energy is estimated [33] as $V_1^{(N)} \sim$ 120
MeV, our estimate for $V_1^{(\Sigma)}$, based on fitting
$\Sigma^-$ atomic Pb, is $V_1^{(\Sigma)} \sim$ 80 MeV.
This value may be compared with the value $V_1^{(\Sigma)} \sim$
60 MeV [20], or $V_1^{(\Sigma)} \sim$ 55 MeV [32], both
estimates following Model D of the Nijmegen group [34] in
agreement with the estimate derived from the phenomenology
of ${^{12}\rm C} (K^-, \pi^\pm)$ reactions [35].

As in ordinary nuclear physics, the $\Sigma$-nucleus
isovector potential cancels partly (for charged
$\Sigma^{\pm}$) the $\Sigma$ Coulomb potential due to the
nuclear charge distribution. For $\Sigma^+$, we have
checked that the attractive symmetry-energy contribution
due to $V_1^{(\Sigma)}$ generally does not overcome the
repulsive isoscalar contribution plus the repulsive
Coulomb energy, so that it is unlikely to bind a
$\Sigma^+$ in nuclei. For $\Sigma^0$, where no symmetry
energy or Coulomb energy contribute, binding is
precluded by the volume repulsion of the isoscalar
$\Sigma$ nucleus potential. These considerations have to
be modified in very light hypernuclei where symmetry
energy dominates over the Coulomb interaction (plus
other charge-dependent effects due mostly to the mass
differences within the $\Sigma$ charge triplet),
leading to a better description of $\Sigma$ hypernuclei
in terms of isospin eigenstates rather than charged
$\Sigma$ states [35]. The only $\Sigma$ hypernuclear
system for which this dominance has been established by
a detailed calculation [36] is ${^4_\Sigma {\rm He}}$
where the $\Sigma$-nucleus isovector potential,
represented as a Lane term of the form (14), is
responsible for the binding of the $I = 1/2,
0^+$ state. Without this Lane term, ${^4_\Sigma {\rm
He}}$ would be unbound.

For $\Sigma^-$, the attractive Coulomb potential due to
the nuclear charge distribution gives rise to an infinite
set of $\Sigma^-$ atomic states. Some of these states,
however, for high $Z$ nuclear cores, may be called
nuclear states inasmuch as the corresponding
wavefunctions are localized within the nucleus or in its
immediate surface region. The energies and wavefunctions
of these ``nuclear" states generated by $V_C$ can be
approximated by noting that for a uniform charge
distribution of a total charge $Ze$ and radius $R$, $V_C
(r<R)$ is a harmonic oscillator potential

\begin{equation}
{V_C = -V_0 + {1 \over 2} M
\omega^2 r^2 \quad ,}
\end{equation}
\begin{equation}
{V_0 = {3 \over 2} Z \alpha
{{\hbar c}\over{R}} \qquad, \quad \qquad \hbar \omega
= {\sqrt{{2 \over 3}
{{V_0}\over{Mc^2}}}}{{\hbar c}\over{R}}\quad ,}
\end{equation}

\noindent
with a depth $V_0$ that grows as $A^{2/3}$ and an
essentially $A$-independent oscillator spacing $\hbar \omega$,
$\hbar \omega = 3.25 \pm 0.03$ MeV between ${^{40}{\rm Ca}}$
and ${^{208}{\rm Pb}}$ for $R = [(5/3)<r^2>]^{1/2}$
in terms of the nuclear charge {\it rms} radius $<r^2>^{1/2}$.
For example, $R = 4.49$ fm for ${^{40}{\rm Ca}}$ and $R =
7.10$ fm for ${^{208}{\rm Pb}}$. Requiring the {\it rms}
radius of the $\Sigma^-$ harmonic oscillator wavefunction
to be less than or equal to R, one gets a series of
nuclear radius values $R_N$

\begin{equation}
{R_0 = 3.87~{\rm fm}~,~~R_1 =
5.00~{\rm fm}~,~~R_2 = 5.92~{\rm
fm}~,~~R_3 = 6.71~{\rm fm}~,~~R_4 =
7.42~{\rm fm}~,}
\end{equation}

\noindent
such that for $R_N < R$, the states of the first
major shells up to and including $N$ are ``nuclear". For
${^{208}{\rm Pb}}$, this suggest that these $\Sigma$
harmonic oscillator states belonging to $N =
0,1,2,3$ are ``nuclear". This is confirmed by a
numerical evaluation using the precise $V_C$, yielding
the following binding energies

\begin{equation}
{{\rm 1s}: 20.35~{\rm MeV}}
\end{equation}
\begin{equation}
{{\rm 1p}: 16.92~{\rm MeV}}
\end{equation}
\begin{equation}
{{\rm 1d}: 13.63~{\rm MeV} \quad ,
\qquad {\rm 2s}: 13.83~{\rm MeV}}
\end{equation}
\begin{equation}
{{\rm 1f}: 10.58~{\rm MeV} \quad ,
\qquad {\rm 2p}: 11.11~{\rm MeV}}
\end{equation}

\noindent
Including $V_{opt}^\Sigma$ (L with $\alpha_\omega =
2/3$) in the binding-energy calculation, we find that
these levels are displaced by 2 to 7.5 MeV to lower
binding energies and also acquire strong-interaction
widths. The 1s level is very broad, about 36 MeV,
whereas the widths of the other levels are relatively
small and vary between 1.2 MeV to 2.6 MeV. The
wavefunctions of these other levels are ``pushed" by
$V_{opt}^\Sigma$ outside the nucleus. These states may be
considered a special case of ``Coulomb assisted states",
a concept put forward in ref. [36]. A reaction which
would excite preferentially a single $\ell$ value for
$\Sigma^-$ in ${^{208}{\rm Pb}}$, could locate these
``nuclear" states for $\ell = 1, 2, 3$. Their location
will give a valuable information on the nature
(attractive or repulsive) of $V_{opt}^\Sigma$ in this
mass region. However, since these states are ``pushed"
outside the nucleus, the cross sections to excite them
by a nuclear reaction are not expected to be sizable.
The chances of establishing a meaningful $\Sigma$
hypernuclear spectroscopy, therefore, are not
particularly encouraging at present.

\vspace*{10mm}
{\it Acknowledgment:}
We wish to thank C.\ J.\ Batty for providing the SP densities used in this
work.
B.K.J. would like to thank the Natural Sciences and Engineering Research
Council of Canada for financial support.

\newpage
\begin{table}
{\bf Table 1 --}
Parameters of DD $\Sigma^-$ nucleus potentials. All parameters
(except $\alpha$) are in units
of fm. Underlined parameters were held fixed during the fits.
\begin{center}
\begin{tabular}{c|c c c c c c c c c} 
 & & & & & & & & & \\
potential & Re$b_0$ & Im$b_0$ & Re$b_1$ & Re$B_0$ & Im$B_0$ & Re$B_1$ &
      $\alpha $ & $\chi^2$ & $\chi^2$ /F \\ \hline
 & & & & & & & & &  \\
  I & 2.1 & 0.50 & $-$1.0 & $-$4.3 &
\underline{0}  & \underline{0} & \underline{0.5} & 16.7 & 0.9 \\
    & $ \pm $ 0.8 & $ \pm $ 0.12 & $ \pm $ 0.4 & $ \pm $ 2.7  &  &  &  &
     &  \\
 & & & & & & & & &  \\
 II & \underline{1.2} & \underline{0.45} & \underline{$-$0.45} &
   $-$2.4 & $-$0.18 & $-$1.9 & \underline{0.8} & 17.8 & 0.9 \\
 & &  &  & $ \pm$ 0.4 & $ \pm $ 0.19 & $ \pm $ 1.4 &  &  &   \\
 & & & & & & & & &  \\
\end{tabular}
\end{center}
\end{table}

\begin{table}
{\bf Table 2 --}
Parameters of the RMF $\Sigma^-$ nucleus optical potentials. The linear
parametrization from ref. [18] was used for the nucleonic sector.
$\alpha_{i} = g_{i\Sigma}/g_{iN}$ ($i=\sigma$, $\omega$, $\rho$), and
$t$ is defined in eq.(13).
\begin{center}
\begin{tabular}{c|c c c c c} 
 & & & & &  \\
potential & $\alpha_{\omega}$ & $\alpha_{\sigma}$
& $\alpha_{\rho}$ & $t$ [MeV fm$^3$] & $\chi^2$ \\ \hline
 & & & & &   \\
 L1 & 1.0 & 0.770 & 2/3 & -400 & 18.1 \\
 & & & & & \\
 L2& 2/3 & 0.544 & 2/3 & -300 & 23.9 \\
 & & & & & \\
 L3 & 1/3 & 0.313 & 0.65 & -265 & 26.5 \\
 & & & & & \\
\end{tabular}
\end{center}
\end{table}

\section*{Figure captions}

\begin{description}
\item[Fig.1: ]Total $\chi^2$ values as a function of the relative notch depth
$d$ (cf. eq.3) superimposed on the best-fit DD optical potential I from
Table 1 with $\alpha=0.5$ (cf. eq.4), for several positions $R_N$
($R_N = R_0 + a_0 \Delta$) expressed via steps $\Delta$ of a nuclear
diffuseness, with a radius $R_0 = 1.1 A^{1/3}$~fm and a diffuseness
$a_0 = 0.5$~fm.

\item[Fig.2: ]Re$V^{\Sigma}_{opt}$ (solid lines) and Im$V^{\Sigma}_{opt}$
(dashed lines) as functions of r for the best-fit DD optical potential I
(cf. eq.4 and Table 1) for Si (Fig.~2a), Ca (Fig.~2b) and Pb (Fig.~2c).
Arrows indicate the position of the corresponding nuclear rms radius.

\item[Fig.3: ]Re$V^{\Sigma}_{opt}$ (solid lines) and Im$V^{\Sigma}_{opt}$
(dashed lines) as functions of r for the RMF $\Sigma^-$ optical potential
NL in Si. Potentials in Figs.~3a, 3b and 3c correspond to the vector meson
coupling ratios $\alpha_{\omega}=1/3$, 2/3 and 1, respectively.

\item[Fig.4: ]Comparison of the linear (L) (Fig.~4a) and nonlinear (NL)
(Fig.~4b) RMF $\Sigma^-$ optical potentials in Pb for the isoscalar vector
meson coupling ratio $\alpha_{\omega}=2/3$. The isovector vector meson
coupling ratio is $\alpha_{\rho}=2/3$ in both cases.

\end{description}
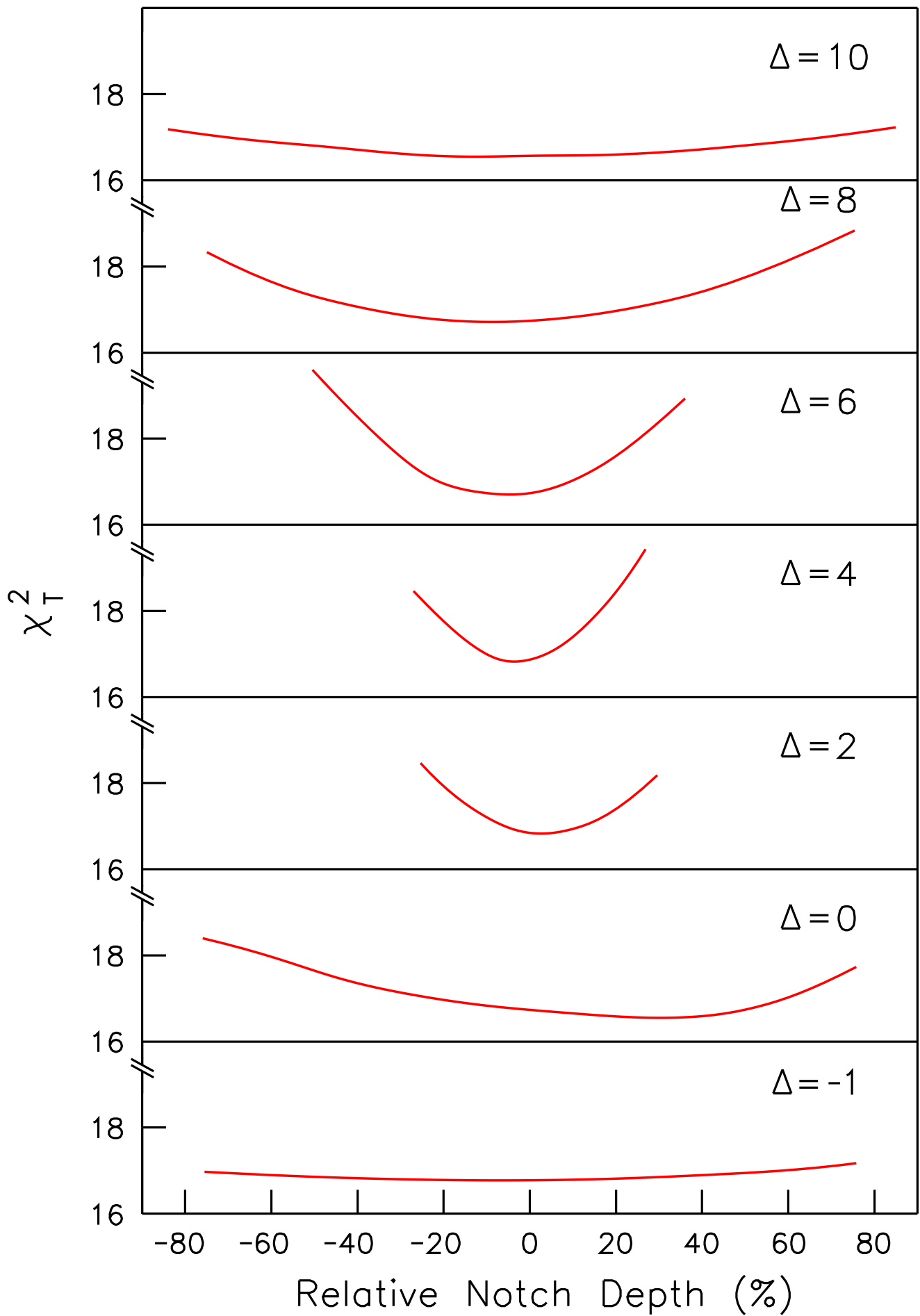
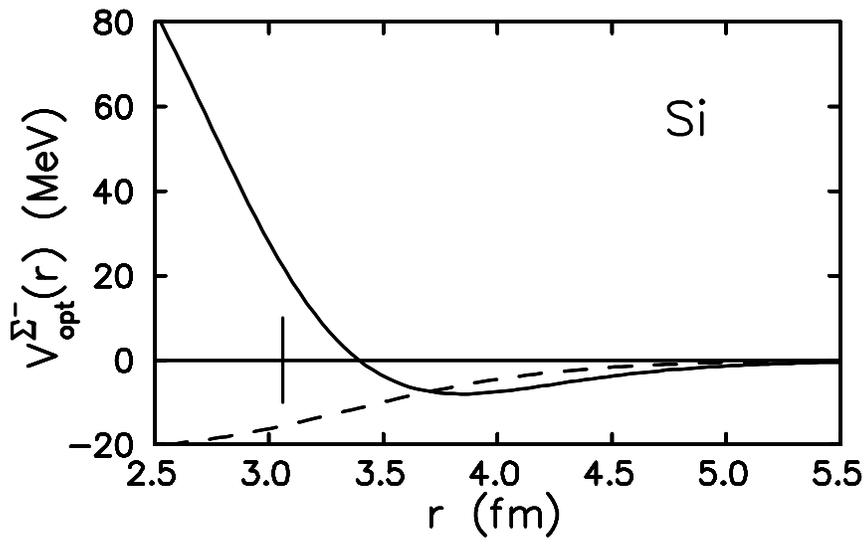
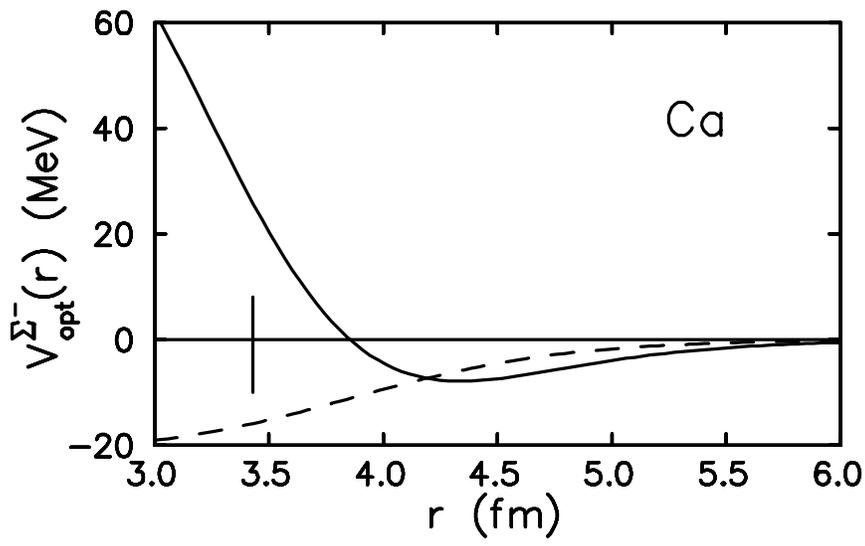
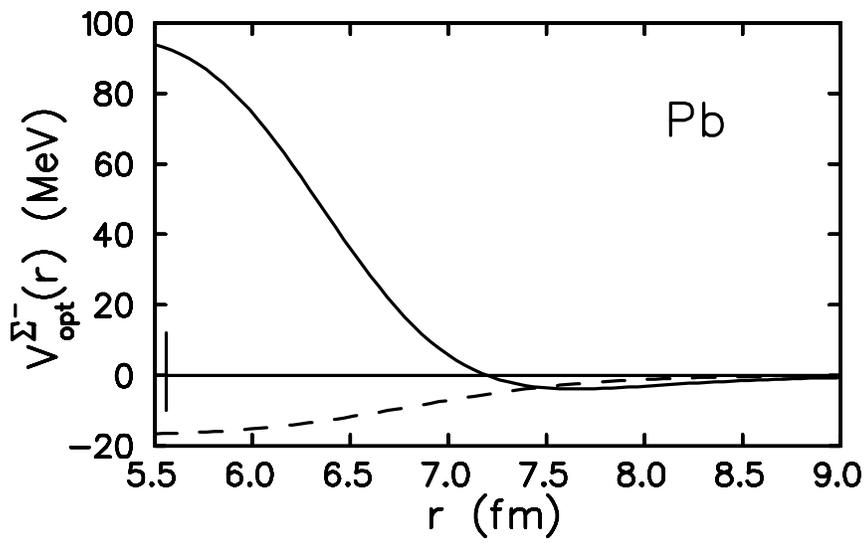
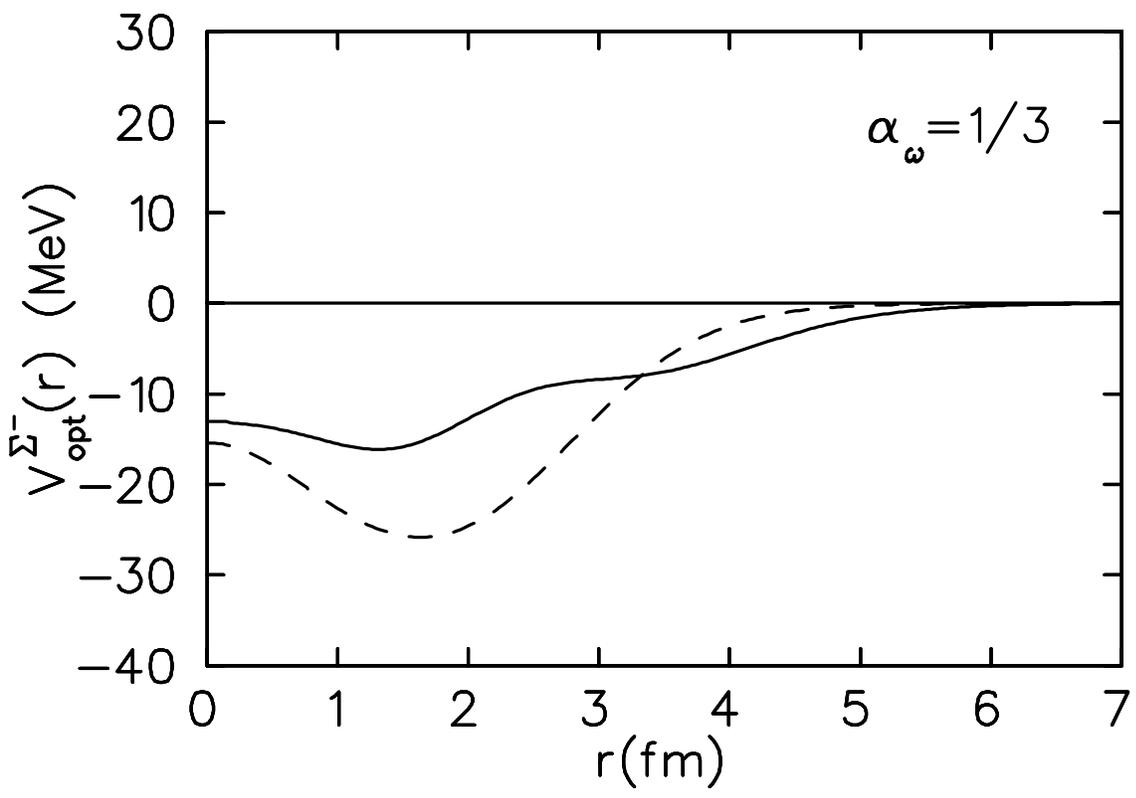
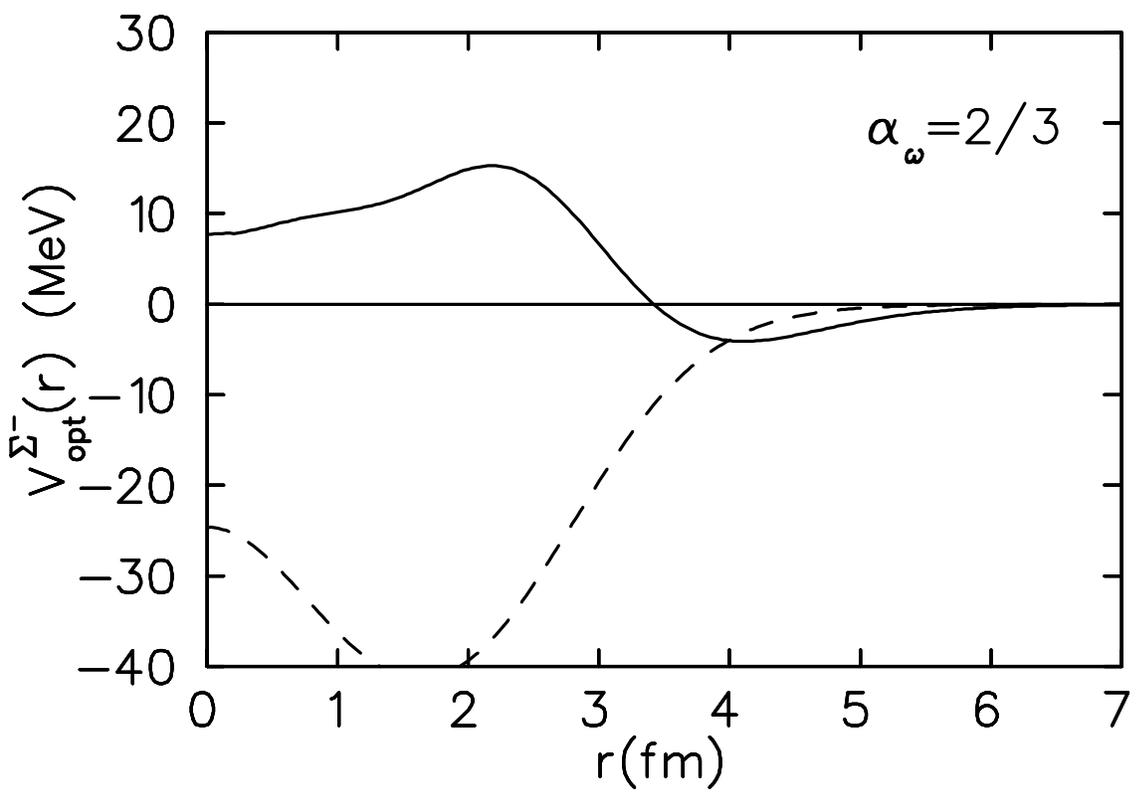
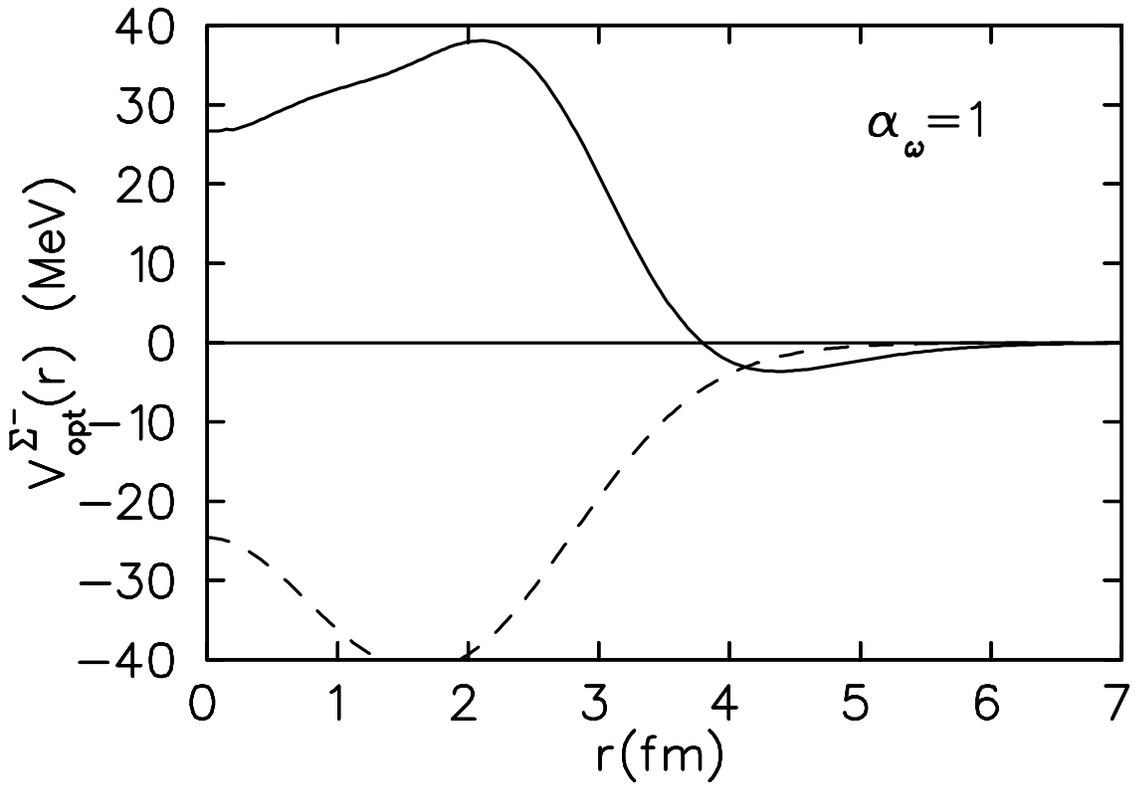
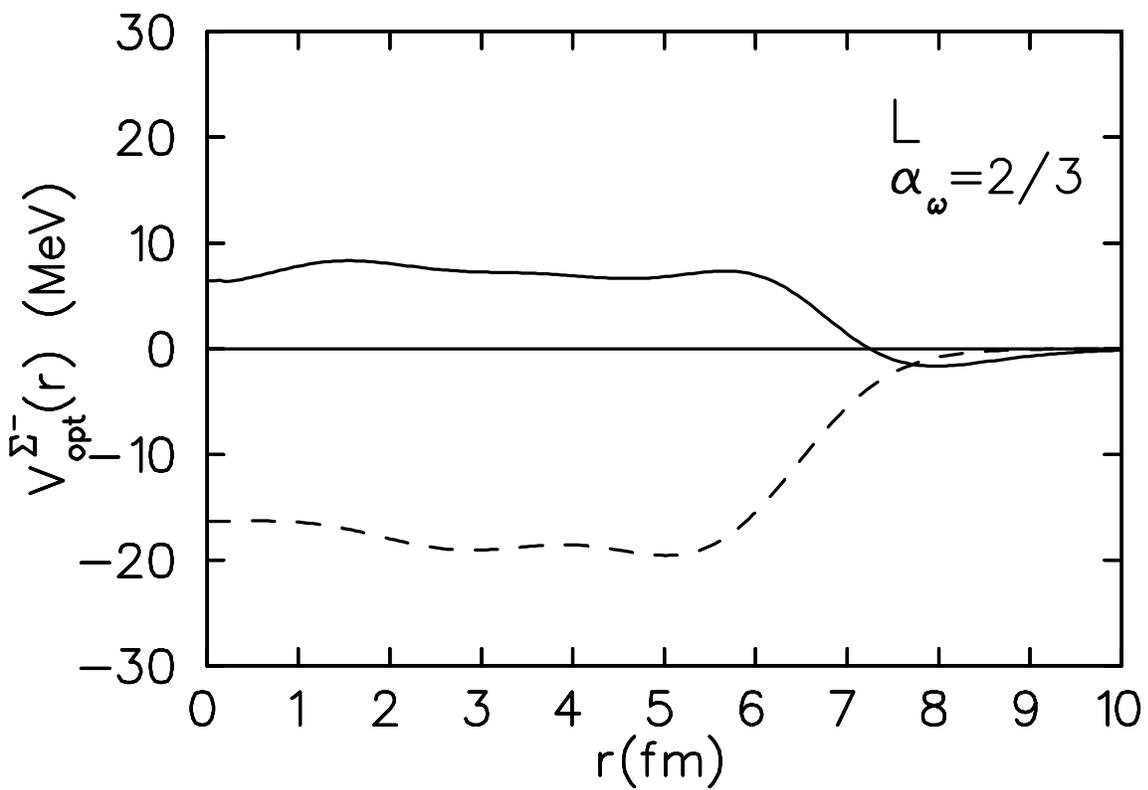
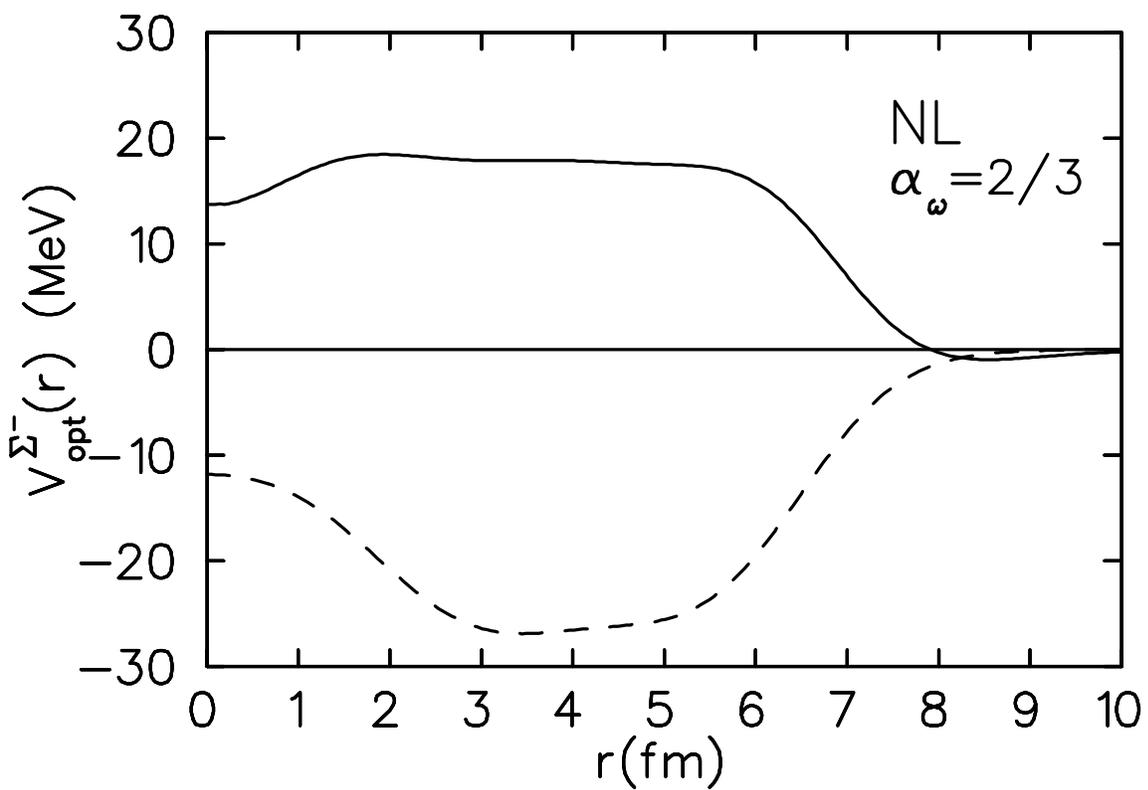
\end{document}